\documentclass[12pt]{article}

\usepackage{scicite}

\usepackage{times}
\usepackage{txfonts} 

\usepackage{graphicx}
\usepackage{bm}

\newenvironment{sciabstract}{%
\begin{quote} \bf}
{\end{quote}}

\newcounter{lastnote}
\newenvironment{scilastnote}{%
\setcounter{lastnote}{\value{enumiv}}%
\addtocounter{lastnote}{+1}%
\begin{list}%
{\arabic{lastnote}.}
{\setlength{\leftmargin}{.22in}}
{\setlength{\labelsep}{.5em}}}
{\end{list}}

\title{An atomic clock with ${10^{-18}}$ instability}

\author
{N.~Hinkley,$^{1,2}$ J.~A.~Sherman,$^{1}$ N.~B.~Phillips,$^{1}$ M.~Schioppo,$^{1}$\\
 N.~D.~Lemke,$^{1}$ K.~Beloy,$^{1}$ M. Pizzocaro,$^{1,3,4}$ C.~W.~Oates,$^{1}$ A.~D.~Ludlow$^{1\ast}$\\
\\
\normalsize{$^{1}$National Institute of Standards and Technology, 325 Broadway, Boulder, Colorado 80305, USA}\\
\normalsize{$^{2}$University of Colorado, Department of Physics, Boulder, Colorado 80309, USA}\\
\normalsize{$^{3}$Instituto Nazionale di Ricerca Metrologica, Str. delle Cacce 91, 10135 Torino, Italy}\\
\normalsize{$^{4}$Politecnico di Torino, Corso duca degli Abruzzi 24, 10125 Torino, Italy}\\
\normalsize{$^\ast$Corresponding author. E-mail: ludlow@boulder.nist.gov}
}

\date{}

\begin{document}


\baselineskip24pt


\maketitle


\begin{sciabstract}

Atomic clocks have been transformational in science and technology, leading to innovations such as global positioning, advanced communications, and tests of fundamental constant variation. Next-generation optical atomic clocks can extend the capability of these timekeepers, where researchers have long aspired toward measurement precision at 1 part in $\bm{10^{18}}$. This milestone will enable a second revolution of new timing applications such as relativistic geodesy, enhanced Earth- and space-based navigation and telescopy, and new tests on physics beyond the Standard Model. Here, we describe the development and operation of two optical lattice clocks, both utilizing spin-polarized, ultracold atomic ytterbium. A measurement comparing these systems demonstrates an unprecedented atomic clock instability of $\bm{1.6\times 10^{-18}}$ after only $\bm{7}$ hours of averaging.

\end{sciabstract}


Periodic physical systems are the basis with which we establish standards of time and frequency. Quantum mechanical absorbers like atoms serve as the best available time and frequency references: they are isolatable, possess well-defined transition frequencies, and exist in abundant identical copies. With over 50 years of development, clocks based on microwave oscillators matched to atomic transitions now define the SI \emph{second} and play central roles in network synchronization, global positioning systems, and tests of fundamental physics \cite{Bauch03a,Levine99}. Under development worldwide, optical clocks oscillate $10^5$ times faster than their microwave predecessors, dividing time into finer intervals \cite{Diddams04a}. While microwave clocks, like Cs fountains, have demonstrated time and frequency measurements of a few parts in $10^{16}$ \cite{Guena2012,Parker2010}, optical clocks are now measuring 1 part in $10^{17}$ \cite{RosenbandHume2008,ChouHume2010,nicholson2012comparison,Gurov13} and have long aspired towards levels of 1 part in $10^{18}$.  Here we demonstrate an optical clock with measurement precision of $1.6$ parts in $10^{18}$.

A measurement at the $10^{-18}$ fractional level is equivalent to specifying the age of the known universe to a precision of less than one second or Earth's diameter to less than the width of an atom. To illustrate the power of this measurement capability, we first consider the gravitational redshift, a consequence of general relativity dictating that clocks `tick' more slowly in stronger gravitational fields. This phenomenon has long been accounted for when remotely comparing atomic clocks with gravitational elevations differing by many meters or km. Instead, a clock with $10^{-18}$ performance can discern changes in the gravitational geoid corresponding to 1 cm of elevation at Earth's surface. It can thus be exploited as a unique and sensitive probe of spatial and temporal variations in the gravitational field \cite{schiller2007optical,chou2010optical,Kleppner2006,Schiller09}, revolutionizing geodesic measurements and potentially impacting areas like hydrology, geology, and climate change studies. Furthermore, alternative gravitational theories predict deviations to the gravitational redshift described by general relativity. Optical clocks at the $10^{-18}$ measurement level, proposed for future space missions, can explore these deviations with more than three orders of magnitude higher precision than presently possible \cite{Schiller09}, as well as stringently test the universality of the gravitational redshift (a consequence of local positional invariance) \cite{schiller2007optical}. Additionally, some unification theories employing non-metric couplings of matter predict variations of the fundamental constants of nature throughout space and time. While these variations are thought to be small, the sensitivity of atomic clocks to these constants (e.g.,~the fine structure constant) already provides a useful constraint on possible variations with time or gravitational field \cite{Chiba2011,RosenbandHume2008,Blatt08a}. A clock with $10^{-18}$ measurement capability offers up to two orders of magnitude tighter constraint on these variations and the theories predicting them, and may even provide insight into dark energy \cite{Fritzsch2012}. Finally, improved timekeepers directly benefit other applications such as navigation systems, synchronization of telescope arrays (e.g.,~VLBI), secure communication, interferometry, and possible redefinition of the SI second \cite{Gurov13}.

To illustrate an essential property of time and frequency standards, imagine a pendulum clock. The periodicity of the pendulum swing provides the clock time base. How this period (or the inverse, frequency) fluctuates over time is a measure of the clock's  \emph{instability}, a characteristic of an oscillator generally quantified by the Allan deviation \cite{GreenhallHowe2011}. No time or frequency standard can make a measurement better than the statistical precision set by its instability. Further, the systematic uncertainty of an oscillator's frequency is often constrained by its long term instability (e.g., slow changes in temperature and air pressure which perturb the pendulum swing). For these reasons, and because many timing applications (including most mentioned above) require only exquisite instability, the instability represents one---if not the most---important property of an atomic standard.

In the pursuit of lower instability, one approach is particularly promising: the optical lattice clock \cite{KatoriTakamoto2003}. In such a clock, a stabilized laser is referenced to ultracold alkaline-earth (or similar) atoms confined in an optical standing wave (i.e., lattice) created by a retro-reflected high-power laser beam. Alignment of the clock interrogation laser along the direction of tight lattice confinement eliminates most Doppler and motional effects while probing the ultra-narrowband electronic `clock' transition. Although the optical lattice induces a Stark shift on the atoms' electronic states, the net effect can be nearly canceled by operation at the so-called `magic' wavelength, $\lambda_m$, where both electronic states of the clock transition are shifted equally \cite{YeKimble2008,KatoriTakamoto2003}. A key advantage of the optical lattice is that many ($10^{3}$ to $10^{6}$) atoms are confined in the lattice potential. All of these atoms are interrogated simultaneously, thereby improving the atomic detection signal-to-noise and thus the instability, which is limited fundamentally by quantum projection noise (QPN) \cite{Itano93a}. Because lattice clocks combine a large atom number with the ultra-narrow optical transition, it has long been hoped that they could realize $10^{-18}$ time and frequency measurement.

The first lattice clocks achieved $10^{-14}$ to $10^{-15}$ level performance \cite{Ludlow06a,LeTargatBaillard,TakamotoHong}, and work has since progressed rapidly worldwide. An important milestone was the demonstration of a lattice clock with short-term (1~s)  frequency instability in the low $10^{-15}$ levels, averaging down to low $10^{-16}$ levels at many hundreds of seconds \cite{LudlowZelevinsky2008}---competitive with the best clock results over similar timescales. Achievement of the lattice clock's full potential was hindered by a technical noise source known as the Dick effect, which arises when an oscillator is only periodically observed \cite{Santarelli98a,DickPrestage}. (In the pendulum clock analogy, imagine trying to discern the pendulum frequency while the room lights flash on and off.) One approach to overcoming this limitation---improving the ultra-stabilized laser used to interrogate the clock transition---demonstrated clock instability well below $10^{-15}$ at short times \cite{JiangLudlow2011}. Alternatively, synchronized interrogation of two lattice clocks with the same stabilized laser exploited rejection of the Dick effect in common mode to probe performance beyond the Dick limit. This technique produced a correlated measurement instability of $4\times10^{-16}$ at short times, approaching $1\times10^{-17}$ for averaging times greater than 1,000~s \cite{TakamotoTakano2011}. Recently, an uncorrelated comparison of two strontium lattice clocks revealed clock instability of $3\times10^{-16}$ at short times, averaging down to $1\times10^{-17}$ in 1,000~s \cite{nicholson2012comparison} or, in another case, reaching the $10^{-17}$ level in 20,000~s \cite{Gurov13}. Here, by comparing two independent optical lattice clocks using ultracold $^{171}\mathrm{Yb}$, we demonstrate an unprecedented clock instability of $1.6\times10^{-18}$ in 25,000~s, a key milestone for applications that have long sought this level of performance.

Both of the Yb lattice clock systems, referred to here as Yb-1 and Yb-2, cool and collect $^{171}\mathrm{Yb}$ atoms from a thermal beam into magneto-optical traps (see Fig.~1). Two stages of laser cooling, first on the strong $^1\!S_0$-$^1\!P_1$ cycling transition at 399~nm, followed by the weaker $^1\!S_0$-$^3\!P_1$ intercombination transition at 556~nm, reduce the atomic temperature from 800~K to 10~$\mu$K. Each cold atom sample is then loaded into an optical lattice with $\sim\!\!300~E_r$ trap depth (recoil energy $E_r/k_B=100$~nK) formed by retro-reflecting approximately 600 mW of laser power, fixed at $\lambda_m\approx759$~nm \cite{LemkeLudlow2009} by a reference cavity. For the measurements described here, about 5,000 atoms captured by each lattice are then optically pumped to one of the $m_{F}=\pm1/2$ ground states using the $^1\!S_0$-$^3\!P_1$ transition. After this state preparation, applying a 140 ms long $\pi$-pulse of 578 nm light resonant with the $^1\!S_0$-$^3\!P_0$ clock transition yields the spectroscopic lineshape shown in Fig.~1C, with a Fourier-limited linewidth of 6 Hz. Experimental clock cycles alternately interrogate  both $m_{F}$ spin states for cancelation of first order Zeeman and vector Stark shifts. The optical local oscillator (LO) is an ultra-stable laser source servo-locked to a high-finesse optical cavity \cite{JiangLudlow2011} and is shared by both Yb systems. Light is frequency-shifted into resonance with the clock transition of each atomic system by independent acousto-optic modulators (AOMs). Resonance is detected by monitoring the ${}^1\!S_0$ ground state population ($N_g$) and ${}^3\!P_0$ excited state population ($N_e$). A laser cycles ground state atoms on the $^1\!S_0$-$^1\!P_1$ transition, and a photomultiplier tube (PMT) collects the resulting fluorescence, giving a measure of $N_g$. After 5-10 ms of cycling, these atoms are laser-heated out of the lattice. At this point, $N_e$ is optically pumped to the lowest-lying $^3\!D_1$ state using a 1388 nm laser. These atoms predominantly decay back to the ground state, where the $^1\!S_0$-$^1\!P_1$ transition is cycled again, now measuring $N_e$. Combining these measurements yields a normalized excitation for the atomic ensemble $N_e/(N_e+N_g)$. During operation of the clock systems, special attention was paid to eliminating residual Stark shifts stemming from amplified spontaneous emission of the lattice lasers, to eliminating residual Doppler effects from mechanical vibrations of the apparatus correlated with the experimental cycle, and to controlling the cold collision shift due to atomic interactions within the lattice \cite{Lemke11a}.

By measuring the normalized excitation while modulating the clock laser frequency by $\pm 3$~Hz, an error signal is computed for each Yb system.  Subsequently, independent microprocessors provide a digital frequency correction $f_{1,2}(t)$ to their respective AOMs, thereby maintaining resonance on the line center. In this way, though derived from the same LO, the individual laser frequencies for Yb-1 and Yb-2 are decoupled, and are instead determined by their respective atomic samples (for all but the shortest time scales). Computers record the frequency correction signals $f_{1,2}(t)$, as shown in Fig.~2A for a 5,000~s interval. Because the experimental cycles for each clock system are not synchronized and have different durations, the recorded correction frequencies are interpolated to a common time base and then subtracted to compute the frequency difference between Yb-1 and Yb-2, as shown in Fig.~2B,C,D. Measurements such as these were repeated several times for intervals of $\sim $15,000~s, demonstrating a clock instability reaching $4 \times 10^{-18}$ at 7,500~s. While collecting data over a continuous 90,000~s interval, we observed the instability curve in Fig.~3, shown here as the total Allan deviation for a single Yb clock. Prior to data analysis, approximately 25\% of the attempted measurement time was excluded due to laser unlocks and auxiliary servo failures. Each servo used to stabilize the laser to the clock transition had an attack time of a few seconds, evidenced by the instability bump near 3~s.  At $\tau=1-5$~s, the instability is comparable to previous measurements of the free-running laser system, and at longer times the instability averages down like white frequency noise as $\sim\!\!3.2 \times 10^{-16} /\sqrt{\tau}$ (for averaging times $\tau$ in seconds), reaching the unprecedented instability of $1.6\times10^{-18}$ at 25,000 seconds.

Also shown in Fig.~3 is an estimate of the combined instability contribution (blue dashed) from the Dick effect and QPN, with the shaded region denoting the uncertainty in these estimates. As can be seen, the observed instability lies close to the combined contributions. We anticipate that significant reductions are possible in the QPN limit by simply using higher atom numbers and longer interrogation times. However, despite earlier reductions in the Dick effect from improved local oscillators \cite{JiangLudlow2011}, Dick noise continues to be an important limit in the performance of this clock.  Looking to the future, this limitation must be reduced so that $10^{-18}$ measurement instability can be realized in $100$ seconds or less. Further stabilization of the optical LO will continue to reduce the Dick effect, both by lowering the laser frequency noise, which is down-converted in the Dick process, and by allowing increased spectroscopy times and thus higher duty cycles. Such laser systems will use optical cavities exhibiting reduced Brownian thermal-mechanical noise by exploiting cryogenic operation \cite{Kessler2012}, crystalline optical coatings \cite{Cole2013}, longer cavities, or other techniques \cite{JiangLudlow2011}.  Fig.~4 demonstrates the benefit of using an optical LO improved over that used in this work, with four times less laser frequency noise and with four times longer interrogation time (corresponding to a short-term laser instability $\sim\!\!5\times10^{-17}$).  The red dotted line gives the Dick instability, while the black dashed line indicates the QPN limit with the same interrogation time, assuming a moderate atomic population of 50,000.

Noting that the calculated Dick effect remains several times higher than the QPN limit, we consider an alternative idea first proposed for microwave ion clocks: interleaved interrogation of two atomic systems~\cite{DickPrestage}. By monitoring the LO laser frequency at all times with two interleaved atomic systems, the aliasing problem at the heart of the Dick effect can be highly suppressed. The solid blue line in Fig.~4 illustrates the potential of a simple interleaved-clock interrogation using Ramsey spectroscopy, which could be even further improved using a more selective interleaving scheme. Even with LO noise levels unimproved from the present work, the Dick effect lies well below a much improved QPN limit (black dashed line). In this case, spin squeezing of the atomic sample could reduce the final instability beyond the standard quantum limit set by QPN (e.g.,~\cite{Leroux2010}).  The two-system, interleaved technique requires spectroscopy on each atomic system to last one half or more of the total experimental cycle.  By extending the clock spectroscopy time to $>$250~ms, we have achieved a 50\% duty cycle for each Yb system, demonstrating the feasibility of this technique. Duty cycles $\geq\!50\%$ can also be realized with the aid of nondestructive state detection \cite{WestergaardLodewyck2010}.

Another important property of a clock is its accuracy, which results from uncertainty in systematic effects that alter the standard's periodicity from its natural, unperturbed state. In 2009, we completed a systematic analysis of Yb-1 at the $3\times10^{-16}$ uncertainty level \cite{LemkeLudlow2009}. Since then, we reduced the dominant uncertainty due to the blackbody Stark effect by an order of magnitude \cite{sherman2012high}. With its recent construction, Yb-2 has not yet been systematically evaluated. The fact that the instability reaches the $10^{-18}$ level indicates that key systematic effects (e.g., blackbody Stark effect, atomic collisions, lattice light shifts) on each system are being well-controlled over the relevant timescales. For all measurements described here, the mean frequency differences $\langle f_{2}(t)-f_{1}(t)\rangle$ were within the Yb-1 uncertainty at $10^{-16}$. Having demonstrated superior instability, we can now efficiently characterize systematic effects on each system, and anticipate $10^{-17}$-level uncertainty in the very near future. Conversely, because long-term instability is typically limited by systematic drifts, further reduction and control of our systems' uncertainties will likely offer improved instability. With continued progress, we envision that $10^{-18}$ instability at 100~s and long-term instability well below $10^{-18}$ could be achieved.


\bibliography{SciRef}

\bibliographystyle{Science}

\begin{scilastnote}
\item The authors acknowledge DARPA QuASAR, NASA Fundamental Physics, and NIST for financial support, and T. Fortier and S. Diddams for femtosecond optical frequency comb measurements.

\end{scilastnote}

\newpage

\begin{figure}[!h]
\begin{center}
\includegraphics[width=6in]{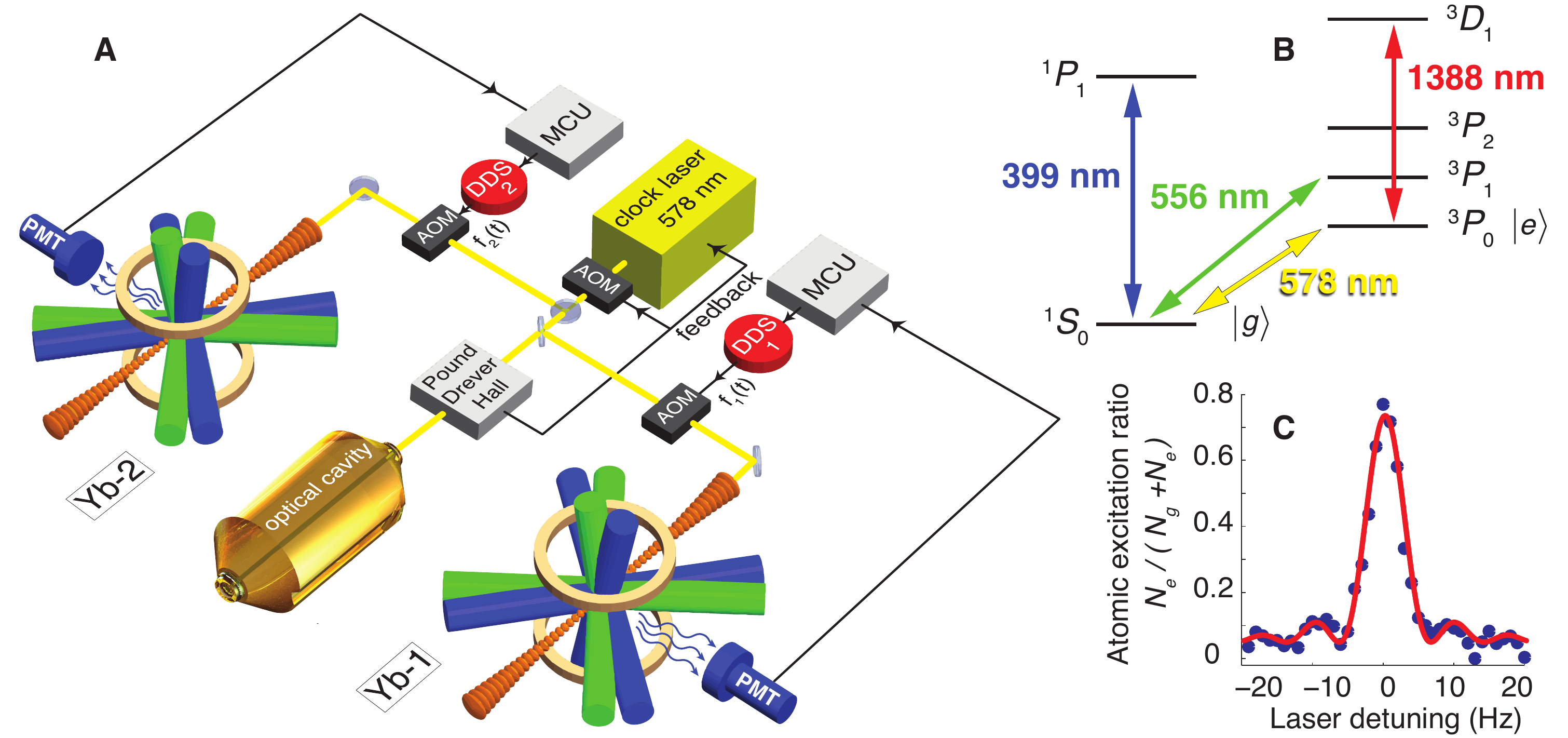}
\label{fig:apparatus}
\caption{\textbf{(A)} Laser light at 578 nm is pre-stabilized to an isolated, high-finesse optical cavity using Pound-Drever-Hall detection and employing electronic feedback to an acousto-optic modulator (AOM) and laser piezoelectric-transducer. This stable laser light is then delivered to the Yb-1 and Yb-2 systems, where it is aligned along the optical lattice axis to probe the atomic clock transition.  Resonance with the atomic transition is detected by observing atomic fluorescence collected onto a photomultiplier tube (PMT).  The fluorescence signal is digitized and processed by a microcontroller unit (MCU), which computes a correction frequency, $f_{1,2}(t)$.  This correction frequency is applied to the relevant AOM by way of a direct digital synthesizer (DDS), and locks the laser frequency onto resonance with the clock transition. \textbf{(B)} Relevant Yb atomic energy levels and transitions, including laser cooling transitions (399 and 556 nm), the clock transition (578 nm), and the optical pumping transition used for excited state detection (1388 nm).  \textbf{(C)} A single-scan, normalized excitation spectrum of the $^1\!S_0$-$^3\!P_0$ clock transition in $^{171}$Yb with 140 ms Rabi spectroscopy time; the red line is a free-parameter sinc$^2$ function fit.}
\end{center}
\end{figure}

\begin{figure}[!h]
\begin{center}
\includegraphics[width=6in]{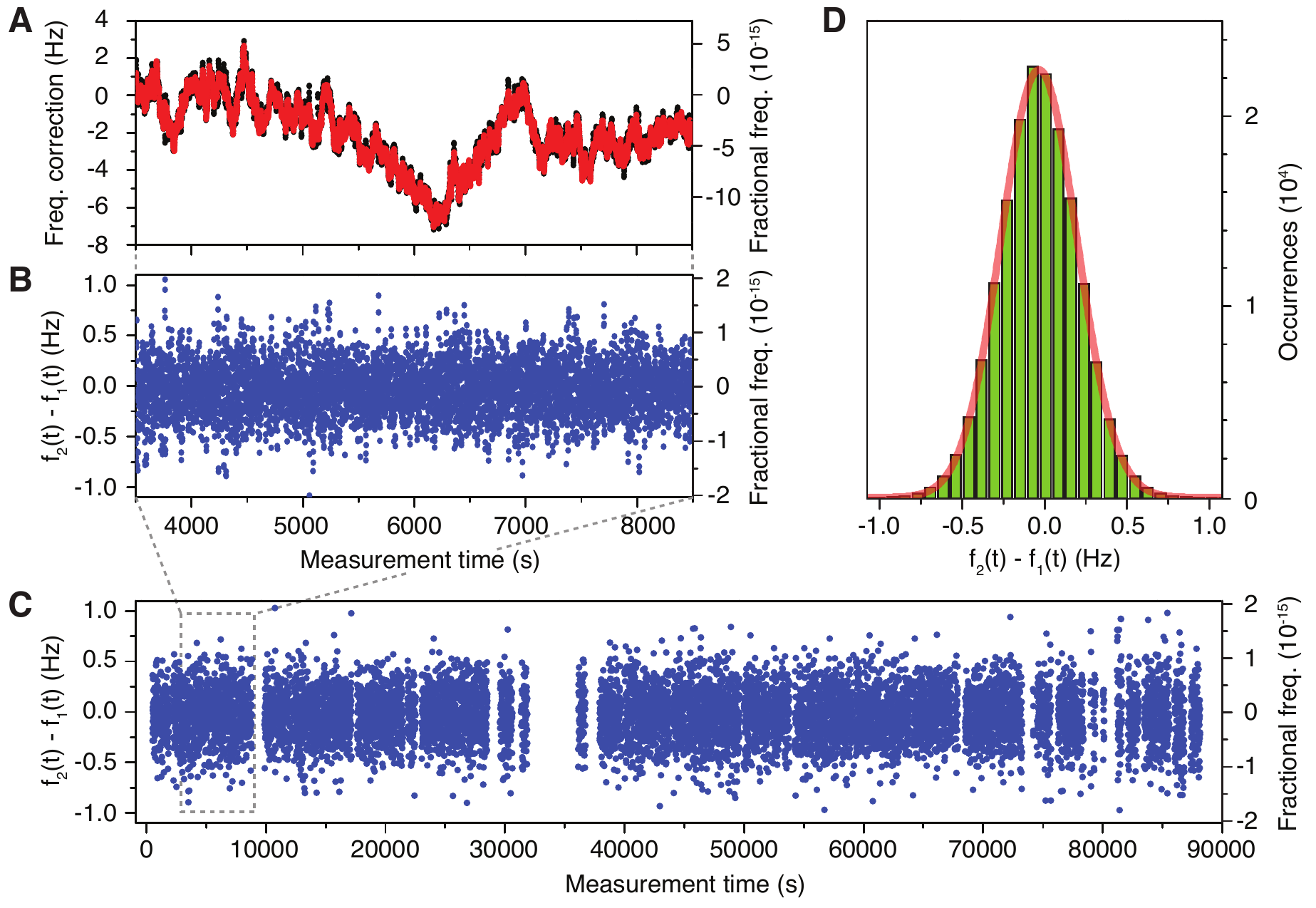}
\label{fig:Allan}
\caption{\textbf{(A)} Correction frequencies, $f_{1,2}(t)$, are shown in red and black. Dominant LO fluctuations are due to the cavity and are thus common to the atomic systems. \textbf{(B)} Frequency difference between the two Yb clock systems, $f_{2}(t)-f_{1}(t)$, for a 5,000~s interval. \textbf{(C)} Data set $f_{2}(t)-f_{1}(t)$ over a 90,000 s interval. Gaps represent data rejected before data analysis due to servo unlocks.   \textbf{(D)} A Gaussian fit of the histogram $f_{2}(t)-f_{1}(t)$ for 90,000~s, with a mean frequency of 30~mHz and $\chi^2_{r}=0.9996$.}

\end{center}
\end{figure}
	
\begin{figure}[!h]
\begin{center}
\includegraphics[width=6in]{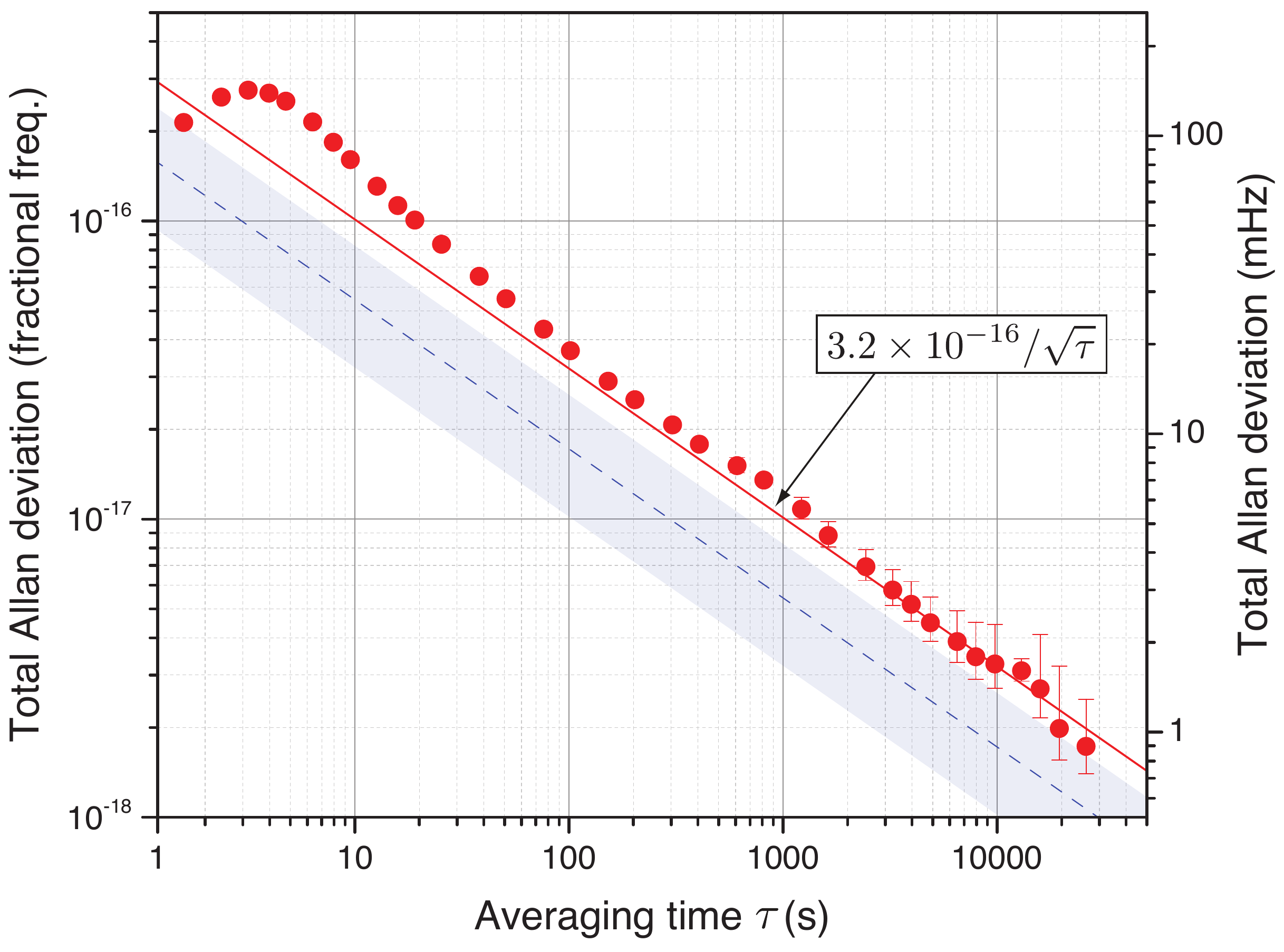}
\label{fig:Dick}
\caption{Total Allan deviation of a single Yb clock, $(f_{2}(t)-f_{1}(t))/\sqrt{2}$ (red circles), and its white-frequency-noise asymptote of $3.2 \times 10^{-16}/\sqrt{\tau}$ (red solid line). The blue dashed line represents the estimated combined instability contribution from the Dick effect ($1.4\times10^{-16}/\sqrt{\tau}$) and QPN ($1\times10^{-16}/\sqrt{\tau}$), with the shaded region denoting uncertainty in these estimates.}
\end{center}
\end{figure}

\begin{figure}[!h]
\begin{center}
\includegraphics[width=6in]{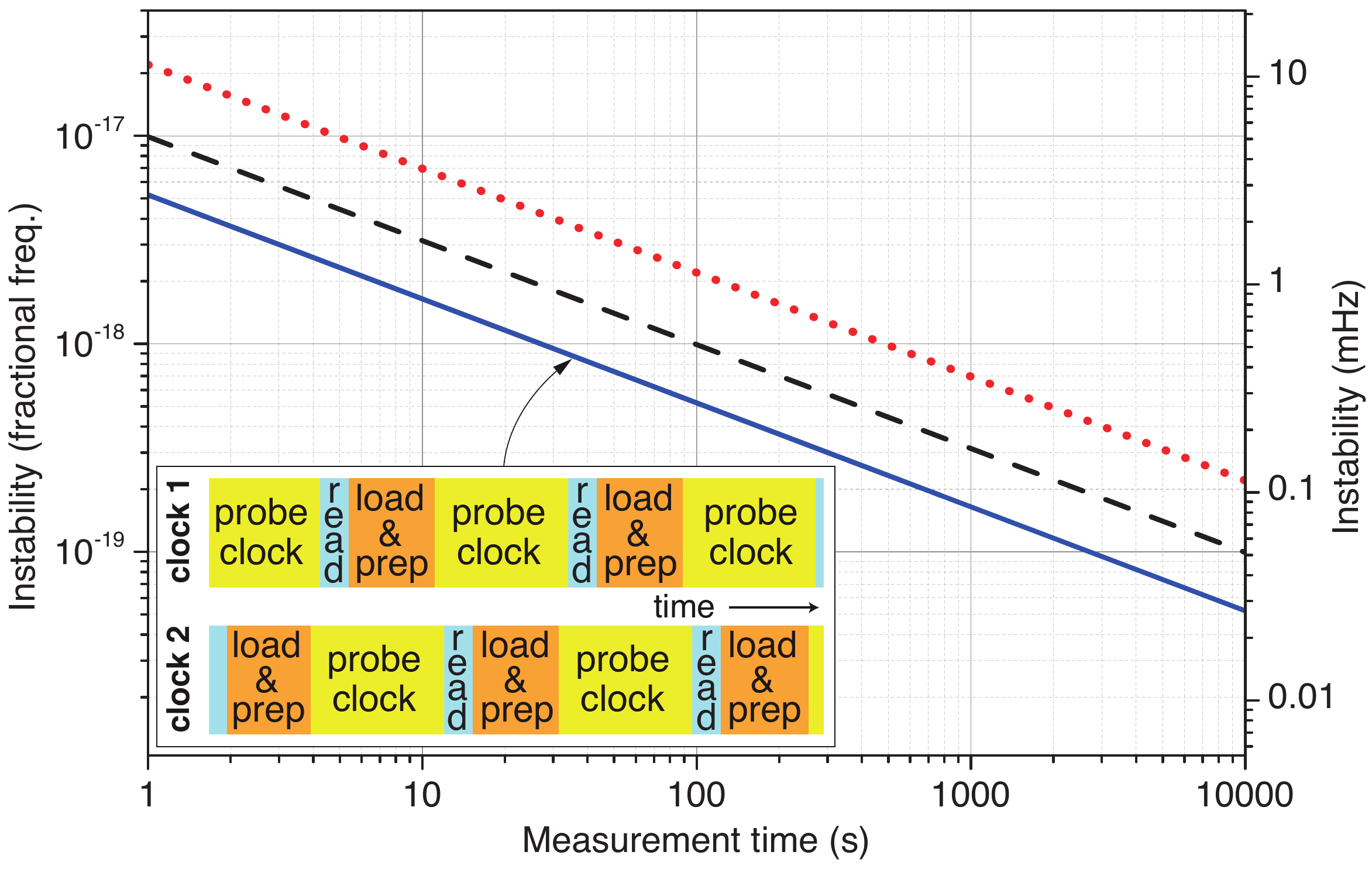}
\label{fig:Dick}
\caption{Calculated instability limits for an improved lattice clock towards the goal of $1\times10^{-18}$ in 100~s. The Dick limit (red dotted) is reduced by using a LO which is four times as stable as that used in this work.  The QPN limit is also shown under the same conditions (black dashed) with a total atom number of 50,000. The inset illustration represents an interleaved interrogation of two atomic systems, allowing continuous monitoring of the LO for suppression of the Dick effect. Dead times from atomic preparation or readout in one system are synchronized with clock interrogation in the second system. The solid blue line indicates the suppressed Dick instability in the interleaved-interrogation scheme of two atomic systems using Ramsey spectroscopy with an unimproved LO. }
\end{center}
\end{figure}





\clearpage
\section*{Supplementary Materials}

\subsection*{Spectroscopy sequences}

\begin{table}[h]
\caption{Spectroscopy sequences for each ytterbium system}
\begin{center}
\label{tab:sequence}
\begin{tabular}{lrr}
							& \multicolumn{2}{c}{\textbf{Duration (ms)}} \\
\multicolumn{1}{c}{\textbf{Step}}	& \multicolumn{1}{r}{\textbf{Yb-1}} & \multicolumn{1}{r}{\textbf{Yb-2}} \\ \hline \hline
1. Blue MOT						& 60		& 74 \\
2. Green MOT stages				& 80		& 80 \\
3. State preparation/Pre-spectroscopy			& 86		& 58 \\
4. 578~nm clock Rabi spectroscopy		& 140	& 140 \\
5. State population readout			& 39		& 43 \\ \hline
\textbf{Total cycle duration $\bm{T_{1,2}}$}	& \textbf{405}	& \textbf{395}
\end{tabular}
\end{center}
\end{table}
Yb-1 and Yb-2 experimental cycles are unsynchronized and have nonidentical but similar durations.  Table~\ref{tab:sequence} details the durations of major operations for both systems.  Initially, a counter-propagating 399~nm beam detuned $\sim\!\!120$~MHz below resonance slows ${}^{171}\!$Yb atoms from a thermal beam.  Unlike many atomic beam magneto-optical traps (MOT)s, we employ no `Zeeman slower'.  Slowed atoms are laser-cooled and confined in a blue MOT consisting of three nearly-orthogonal, retro-reflected 399~nm beams detuned 15~MHz below resonance and a magnetic field gradient of $\sim\!\!3$~mT/cm.  The three green MOT stages, lasting 30 ms, 30 ms, and 20 ms respectively for both systems, use increasingly resonant 556~nm light and a smaller but varying magnetic field gradient to further cool the atoms. Notably, doubling the field gradient during the second stage to $\sim\!\!2$~mT/cm compresses the atomic cloud, ensuring a better overlap with the optical lattice.

After extinguishing the green MOT beams, we apply a $\sim\!\!0.5$~mT magnetic field and spin-polarize trapped atoms by optical pumping on $^1\!S_0$-${}^3\!P_1$.  Another $\sim\!\!0.1$~mT field lifts the Zeeman degeneracy during 578~nm clock spectroscopy. Stabilized 578~nm light is routed to the atomic systems through Doppler-noise-canceled fiber optic paths. An acousto-optic-modulator (AOM) directs 578~nm light pulses onto the atoms.  To suppress phase chirps from AOM switching, we drive this AOM with $<$10~mW of radio-frequency power. Rabi spectroscopy with a 140~ms pulse time provides a convenient combination of narrow lineshape, short clock cycle, and reliability when locking to the clock transition. Detection of atomic excitation following spectroscopy is detailed below. During spectroscopy, all resonant laser beams are extinguished when appropriate by both AOMs and shutters, with the exception of 578~nm light.  The optical lattice beam is continuously applied with actively stabilized intensity.

\subsection*{State detection, normalization, and background subtraction}
As described in the main text, we apply a series of resonant pulses of 399~nm light to count the atomic population via scattered fluorescence.  The combined photon collection efficiency is $<~\!\!\!1\%$ due to small observational solid angle, photo-multiplier tube (PMT) quantum efficiency, and transmission efficiency of a 399~nm bandpass filter.  However, by retroreflecting the probe light to improve the balance of optical forces, each atom scatters $>$$10^3$ photons before being laser-heated out of lattice confinement.  Unfortunately, probe light also illuminates the thermal effusive Yb beam, creating a fluorescence signal that, together with scattered laser light, constitutes an important background source.  Identifying the integrated fluorescence from trapped ground state atoms as $N_g$, and the integrated background signal as $B$, the first readout in the sequence is recorded as $P_1 = N_g + B$.  Following this, a second identical pulse ideally generates an integrated signal $P_2 = B$.  Then, application of resonant 1388~nm light optically pumps nearly all ($>~\!\!\!90\%$) trapped atoms from the ${}^3\!P_0$ state into the ground state through the ${}^3\!D_1$ state.  Finally, a third identical 399~nm probe pulse yields the integrated signal $P_3 \simeq N_e + B$.

We synthesize an atomic excitation fraction $\epsilon$, normalized against total atom number and suppressing background effects, from the three records $P_{1,2,3}$:
\begin{equation}
\epsilon \equiv \frac{N_e}{N_g + N_e} \approx \frac{ P_3 - P_2}{P_1 + P_3 - 2P_2}.
\end{equation}

In order to cleanly implement the atomic state detection, the 399~nm laser is (1) frequency stabilized to the ${}^1\!S_0$-${}^1\!P_1$ transition using a modulation transfer spectrometer, (2) intensity stabilized, (3) polarization filtered, and (4) sufficiently intense to drive the  ${}^1\!S_0$-${}^1\!P_1$ transition into saturation.

\subsection*{Clock difference-signal}
Each Yb atomic servo consists of a micro-controller unit (MCU) updating a direct-digital synthesizer (DDS) with a correction signal $f_{1,2}$.  System Yb-$n$ applies discrete corrections $f_n(t^{(n)}_i)$ at unsynchronized times $t^{(n)}_i$.  Independent computers record and timestamp each correction $f_n(t^{(n)}_i)$.  During post-processing, we employ a piecewise-cubic Hermite interpolating polynomial to establish $f_1$ and $f_2$ on a common set of timestamps separated by the average of the two systems' cycle durations, $\bar{T} = \frac{1}{2} \left( T_{1} + T_{2}  \right)$. As described in the main text, the difference in the correction signals $f_{2}(t)-f_{1}(t)$ corresponds to the frequency difference between the atomic systems, and the instability of $f_{2}(t)-f_{1}(t)$ gives the combined instability measured between the clocks. If systematic variations of each atomic frequency existed and they were positively correlated between the systems, then these variations would be reduced in the frequency difference. In such a case, the measured instability would not reflect the true instability of the lattice clocks.  We note that by design, these atomic systems are largely independent of each other.  Their most significant shared attribute is co-location in the same laboratory.  Thus, while local heat loads independently influence each atomic apparatus, the ambient laboratory temperature can drive temperature correlations that influence the blackbody Stark shift.   However, we note the mean laboratory temperature was long-term stable to $<~\!\!\!50$~mK over the measurement period.  Most key systematic effects, like the cold collision shift, Zeeman shifts, probe light shifts, and other technical effects, are independent for each system.  The optical lattice lasers are independent systems, and their frequencies are separated by 1~MHz, both held close to the magic wavelength ($\lambda_m$) . We note that control of this optical frequency at the MHz level is routine.

\subsection*{Lattice laser stabilization}

Each Yb system's lattice is derived from a distinct Titanium Sapphire (Ti:S) laser that is injection locked by an external cavity diode laser near $\lambda_m\approx759$ nm. A small fraction of the nominal $\sim$1.5~W output power from the Yb-2 lattice system is sent via polarization-maintaining (PM) optical fiber to a reference optical cavity. Feedback to an intra-cavity piezoelectric transducer (PZT) stabilizes the Yb-2 lattice laser frequency to the cavity resonance. An AOM upshifts the laser frequency by $80$~MHz, bringing it to 394,798,295.4~MHz for confinement of Yb-2 atoms near $\lambda_m$.

The frequency difference between the Ti:S of Yb-1 and that of Yb-2 is directly detected in a heterodyne optical interferometer.  This heterodyne beat is stabilized to $\sim\!\!162$~MHz using an RF delay line interferometer, with Yb-1's laser frequency chosen higher than that of Yb-2.  An AOM downshifts Yb-1's laser frequency by $82.3$~MHz, to 394,798,294.3~MHz before the light is sent via PM fiber to the atomic system. AOMs are also used to intensity stabilize each lattice beam.

\end{document}